\documentclass[aps,prd,onecolumn,groupedaddress,showpacs,nofootinbib,amssymb]{revtex4}
\usepackage[dvips]{graphicx}
\usepackage{amssymb}
\usepackage{amsmath}
\usepackage{graphicx,,color}
\usepackage{amsfonts}
\usepackage{bm}
\usepackage{cancel}
\usepackage{comment}
\newcommand\be{\begin{equation}}
\newcommand\ee{\end{equation}}

\allowdisplaybreaks[4]

\begin{document}

\title{Is a Topology Change After a Big Rip Possible?}
\author{V.K. Oikonomou,$^{1,2,3}$\,\thanks{v.k.oikonomou1979@gmail.com}}
\affiliation{
$^{1)}$ Department of Physics, Aristotle University of Thessaloniki, Thessaloniki 54124, Greece\\
$^{2)}$ Laboratory for Theoretical Cosmology, Tomsk State University
of Control Systems
and Radioelectronics, 634050 Tomsk, Russia (TUSUR)\\
$^{3)}$ Tomsk State Pedagogical University, 634061 Tomsk, Russia\\
}

\tolerance=5000

\begin{abstract}
Motivated by condensed matter physical systems, in which a finite-time singularity indicates that the topology of the system changes, we critically examine the possibility of the Universe's topology change at a finite-time cosmological singularity. We emphasize on Big Rip and Type II and IV cosmological singularities, which we classify to future spacelike and timelike singularities. For the Type IV and Type II singularities, since no geodesics incompleteness occurs, no topological change is allowed, by using Geroch's theorem arguments. However, for the Big Rip case, Tipler's arguments allow a topology change, if the spacetime in which the topology change occurs is non-compact and the boundary of this region are two topologically distinct three dimensional spacelike partial Cauchy hypersurfaces. Also, some additional requirements must hold true, among which the weak energy condition, which can be satisfied in a geometric way in the context of a modified gravity. We critically examine Tipler's arguments for the Big Rip case, and we discuss the mathematical implications of such a topological change, with regard to the final hypersurface on which geodesics incompleteness occurs.
\end{abstract}

%PACS numbers: 04.50.Kd, 95.36.+x, 98.80.-k, 98.80.Cq

\maketitle

\section{Introduction}

One of the most peculiar physical to some extent, situations in the classical description of nature is the occurrence of singularities. One of the most common singularities in classical electrodynamics is the singularity of the Coulomb potential at the origin of the source of the potential, that is at $r=0$. Feynman and his description of Quantum Electrodynamics, resolved the singularity of the electron potential, by including quantum self interactions of the electron, and thus explaining that the bare electric change is infinite, but the renormalized one is not. This discussion indirectly implies that singularities in classical physics are in some sense, or could be viewed as flaws of the classical description. Perhaps, however in the physics of condensed systems, the occurrence of singularities of finite-time type in the physical quantities that describe the system, indicate one phenomenological situation which is particularly interesting, topology change \cite{verga,hele}. This is known to occur in Hele-Shaw systems, for example, where the pressure of the surface describing a drop, diverges at some finite-time instance, and the surface of the drop changes topology exactly at the singularity. Also for other statistical mechanics systems, a phase transition is accompanied by topology change in the configuration space \cite{conf}. Motivated by this phenomenon of condensed matter physics, in this paper we speculate and investigate in a critical way, the possibility of having topology change accompanying a finite-time cosmological singularity. At first, we need to note two things: the two situations are utterly different technically, and also that we do not make any assumption on the Universe having a non-trivial topology as a whole. In fact, we are interested in the possibility that the Universe actually changes its topology after a finite-time cosmological singularity occurs, emphasizing on crushing type singularities and specifically on the Big Rip. Big Rip singularities quite frequently occur in modified gravity descriptions of cosmology \cite{Nojiri:2017ncd,Bamba:2008ut,Briscese:2006xu,Nojiri:2006ww}, but can also occur for phantom fluids or fields in classical general relativity \cite{Caldwell:2003vq}. Also, the quantum description of cosmology, like the most promising loop quantum cosmology \cite{LQC1,LQC3,LQC4,LQC5}, is known to eliminate all types of finite-time singularities \cite{Sami:2006wj}. Regardless of this aspect, in this paper we shall investigate the possibility that topology change occurs in a Big Rip singular spacetime, in the context of classical cosmology theories of any type.

The topology change issue in general relativity contexts is vastly addressed in the 70's, and the results are known for quite some time \cite{Geroch:1967fs,Tipler:1986qu,Gibbons:1994pr,Tipler:1977eb,Gibbons:1992he,Gibbons:1992fh,Gibbons:1991tp}. In the seminal paper \cite{Geroch:1967fs}, Geroch gave general conditions under which topology change in general relativity may occur, and the result was that if the spacetime is not singular, or if causality is not violated, quantified in the presence of closed timelike geodesics (CTG), then topology change cannot occur for any Lorentzian four dimensional manifold. Gibbons and Hawking \cite{Gibbons:1992he} further proved that if such a topology change occurred, then it would lead to the problem of non-having fermions in the topologically different spacetime. Thus, this clearly motivates our study to investigate whether topology change can occur after a crushing type event, such as a Big Rip or even other finite-time singularities. This problem was addressed in some detail by Tipler in \cite{Tipler:1977eb} on which we shall base our analysis and we shall discuss the various features of Tipler's theorems for cosmological spacetimes which can be singular at some finite-time. A very important issue for our study is the definition of singularities in cosmology. Penrose and Hawking provided a seminal series of papers regarding the singularities in general relativity \cite{Hawking:1969sw,Hawking:1967ju,Hawking:1966vg,Penrose:1964wq,Penrose:1988ph} and Penrose included in the description of a singularity, the geodesics incompleteness argument. This will be mainly adopted in this work, along with the fact that on singularities, the higher curvature invariants of four dimensional spacetime diverge, as it happens for example on crushing type singularities. In cosmology, the classification of finite-time singularities was firstly performed in \cite{Nojiri:2005sx}, where the total effective energy density and pressure, along with the scale factor (for a Friedmann-Robertson-Walker (FRW) metric) were used as factors for classifying a singularity. As we already noted, for the needs of this study, we shall deal with singularities for which geodesics incompleteness occurs, since this is a vital argument in Tipler's theorems \cite{Tipler:1977eb}, and also the fact that curvature scalars, like the  Kretschmann scalar, diverge at the singularity. In what follows, we shall critically address the conditions that must hold true in order for a topology change to occur after a Big Rip singularity, and also we discuss the same possibility for other cosmological finite-time singularities. We also classify all the known finite-time cosmological singularities to future spacelike and timelike singularities, by using the geodesics incompleteness criterion. We also very briefly discuss the quantum aspects of the problem, which eventually may resolve the Big Rip singularity issue of certain classes of classical cosmology.

\section{Finite-time Singularities Classified to future-Spacelike and Timelike Singularities}

The cosmological finite-time singularities were firstly classified in \cite{Nojiri:2005sx}, by using the scale factor $a(t)$, the total effective pressure $p_{eff}$ and the total effective energy density $\rho_{eff}$, and assuming a homogeneous and isotropic Universe with scale factor $a(t)$. Particularly there are four distinct types of finite time singularities, the following,
\begin{itemize}
\item Type I Singularity (``The Big Rip Singularity''): This is a crushing type singularity, the most severe finite-time singularity from a phenomenological point of view. This type of singularity occurs quite frequently in modified gravity contexts \cite{Nojiri:2017ncd}, and in this case, at the singularity time instance $t=t_s$, the scale factor, the energy density and the pressure of the Universe diverge.
\item Type II Singularity (``The Sudden Singularity''): This type of singularity is merely a pressure singularity, since only the pressure diverges at the finite-time instance $t=t_s$. It is known as pressure singularity \cite{barrowsudden,barrowsudden1}.
\item Type III Singularity: The second most severe singularity from a phenomenological point of view. In this case, the scale factor is finite and only the pressure and the energy density diverge.
\item Type IV Singularity: This is a soft type singularity, in which case all the physical quantities we mentioned earlier are finite and only the higher derivatives of the Hubble rate diverge. Extensive studies on the phenomenology of these singularities can be found in Refs. \cite{Odintsov:2015gba,Odintsov:2015jca,Barrow:2015ora,Nojiri:2015fra,Oikonomou:2015qfh}.
\end{itemize}
Essentially, the finite-time singularities described above, affect the physical quantities that can be defined on a three dimensional spacelike hypersurface that is defined by the time instance $t=\tau_s$, namely the pressure and the energy density, but also affects a geometric related parameter that enters the homogeneous and isotropic metric, namely the scale factor. In effect, only the Big Rip is an actually metric singularity, while the rest affect only the physical quantities on the three dimensional hypersurface $t=\tau_s$.

However it is vital for our analysis to know which singularities lead to geodesics incompleteness at exactly the three dimensional spacelike hypersurface $t=\tau_s$. This task has been performed in Ref. \cite{FernandezJambrina:2004yy}, in the context of an isotropic and homogeneous cosmological evolution, with either flat, hyperbolic or elliptic spatial sections. Particularly, as was shown in Ref. \cite{FernandezJambrina:2004yy}, for the case of Big Rip, geodesics incompleteness and also the higher curvature scalars diverge. To be specific, for the higher curvature scalars, causal geodesics meet a strong singularity at the value $t=\tau_s$ of its  affine parametrization, if the following expression diverges at $t=\tau_s$,
\begin{equation}\label{highercurvsc}
\int_0^{\tau}dt R^{i}_{0j0}(t)\, .
\end{equation}
For the Big Rip and the Type III case, the above diverges and also for the Type II (sudden) and for the Type IV this is finite. However, among the Type III and the Big Rip, only the latter is an actual metric singularity, so the spacetime has an actual singularity at the whole three dimensional spacelike hypersurface, so we focus on this type of singularity in this paper, when we mention a crushing type singularity. The Type III singularity requires some more deeper analysis, plus it is not a metric singularity so we do not include it in our considerations. The Type III case, involves uncontrolled tidal forces at the singularity, which would rip apart extended objects, since it leads to geodesic deviations and in effect to strong tidal forces at $t=\tau_s$.

From a phenomenological point of view, the Big Rip and the Big Bang singularity are the same, since the Big Rip singularity occurs in the whole spatial slice $t=\tau_s$, and in addition it affects both the spacetime quantities, like the scale factor, but also all the physical quantities defined on the spacelike hypersurface $t=\tau_s$. On the other hand, in the Big Bang case, the scale factor is zero, and the geodesics equations cannot be extended before this singularity. Moreover, the initial singularity cannot be considered as a point, as it usually is, since this would imply an infinity of overlapping particle horizons, and in effect, as the evolution begins, this would lead to an infinity of causally disconnected regions in the Universe \cite{Penrose:1988ph}. Thus, following Penrose's description, the Big Bang singularity is merely a singularity at a three dimensional hypersurface, which can be achieved for the $a=0$ one by a conformal rescaling. Regardless of this description, in both the Big Rip and the Big Bang singularities, geodesics incompleteness occurs, in the opposite time directions. Following Penrose's classification of spacelike and timelike singularities, let us classify the Big Rip, the Big Bang and the Type II and Type IV singularities. The Big Rip singularity is a future spacelike singularity, since starting from any acausal partial Cauchy three dimensional hypersurface $\mathcal{S}(\tau)$, each point $p$ belonging in $\mathcal{S}$ has the property that $\bigcup I^{+}(p)=\mathcal{S}(\tau_s)$, where $\mathcal{S}(\tau_s)$ is the singular hypersurface at $t=\tau_s$, for the case of a finite-time singularity. Simply this would mean that the spacetime ends at the singular hypersurface, so a timelike past endless singularity has a future end point. Accordingly, the Type II and Type IV singularities, are timelike singularities, since on these, at the singular hypersurface $\mathcal{S}(\tau_s)$, both past endless and future endless geodesics meet. This is a consequence of the fact that on Type II and Type IV, no geodesics incompleteness occurs. However, we could extend Penrose's classification to introduce a new type of singularity, the intermediate spacelike singularity, in which both past endless and future endless geodesics meet, in an extended spacelike hypersurface.

Finally, as noted by Penrose \cite{Penrose:1988ph}, the Big Bang is a past spacelike singularity. Note above that we made a crucial assumption for the initial hypersurface $\mathcal{S}(\tau)$ to be a partial Cauchy hypersurface. This is very important, since each timelike curve should thread this hypersurface only once. Regarding the future spacelike Big Rip singularity, the fact that a singularity occurs in the spacetime, and it is a metric singularity, means that a Cauchy horizon exists, so $\mathcal{H}^{+}(\mathcal{S}(\tau))\neq \emptyset$. This will be important for the analysis that follows.

\section{Finite-time Singularities and Topology Change: Investigation of the Possibility and Constraints}

Having classified the Big Rip, along with the Type II and IV singularities to spacelike and timelike, in this section we shall investigate the conditions that need to hold true in order for topology change to occur at singular spacetimes. For simplicity we focus on spacetimes that have compact spacelike hypersurfaces, not necessarily positively curved. For the case of Type II and Type IV, no geodesics incompleteness occurs, thus the spacetime metric is not singular. This simply means that for a homogeneous and isotropic Lorentzian metric with compact spatial sections, Geroch's theorem applies, according to which,
\begin{itemize}
    \item If a compact subset $M$ of an isochronous Lorentzian spacetime has no CTG and its boundary is $\partial M=\mathcal{S}\bigcup \mathcal{S}'$, where $\mathcal{S}$ and $\mathcal{S}'$ are compact spacelike hypersurfaces, then the hypersurfaces $\mathcal{S}$ and $\mathcal{S}'$ are diffeomorphic and the topology of $M$ is $S\times [0,1]$.
\end{itemize}
Practically Geroch's theorem in our case ensures that given a compact subspace of a Lorentzian manifold, if this does not contain any singularities, or equivalently if it does not contain any CTG, then topology change cannot occur in this subset $M$. Also the topology of this manifold is locally trivial, since it is $S\times [0,1]$. Hence, since the presence of a singularity means geodesic incompleteness and for the case of Type IV and Type II singularities, no geodesics incompleteness occurs, this means that no topology change occurs for sure for these two types of singularities.

Let us now proceed to the examination of the Big Rip case, in which case, geodesics incompleteness occurs in the spacetime manifold, at some three dimensional hypersurface $t=\tau_s$. For this case we will need to use more formal arguments in order to perfectly describe the situation. We shall use some of Tipler's arguments in order to perfectly describe the situation at hand \cite{Tipler:1977eb}. Firstly, we assume that at some time before the actual singularity, the Universe is described by a three dimensional spacelike hypersurface at $t=\tau_i$, denoted as $\mathcal{S}$, which has all the regular initial data which emerge from $I^{-}(\mathcal{S})$. These data are regular, meaning that no singularity occurs on this surface. Also we assume that the hypersurface $\mathcal{S}$ is a partial Cauchy surface, in order to ensure that every timelike geodesic threads the hypersurface only once. Consider now that there is a time interval $[\tau_i,\tau_f]$, and also that the hypersurface $\mathcal{S}'$ corresponds to the time instance $\tau_f$, and also that these surfaces are themselves compact and these define the disjoint union boundary of a non-compact Lorentzian subset $\mathcal{B}$ of a spacetime $\mathcal{M}$, that is $\partial \mathcal{B}=\mathcal{S}(\tau_i)\bigcup \mathcal{S}(t_f)=\mathcal{S}'$. For this theoretical setting, Tipler's arguments state that (using the notation we used for the partial Cauchy hypersurfaces),

\begin{enumerate}
    \item If the weak energy condition  and the Einstein equations holds true.
    \item If for any causal vector $K^{a}$, the following holds true $K^{a}K^{b}K_{c[}R_{d]ab[e}K_{f]}\neq 0$ at least at one point $p$ on the geodesic on which $K^{\alpha}$ is tangent.
    \item If $d(\mathcal{S}(\tau_i),[\mathrm{int}\mathcal{D}^{+}(\mathcal{S}(\tau_i))\bigcap \mathcal{B})$ is finite.
    \item And finally if $\mathcal{S}(\tau_i)$ and $\mathcal{S}(t_f)=\mathcal{S}'$ are not diffeomorphic.
\end{enumerate}
Then, the spacetime $\mathcal{M}$, to which $\mathcal{B}$ is a non-compact subset,  is timelike geodesically incomplete. Also in Tipler's theorem there is much more, which can be found as surrounding implications of the theorem, so we quote these at this point, starting with the fact that Tipler's theorem proves that the actual topology change occurs in a spacetime which is geodesically incomplete, and that satisfies the requirements given above. Also the topology change occurs in a finite non-compact subset of the spacetime $\mathcal{M}$, namely in $\mathcal{B}$. The word finite should not be confused with the finiteness of a region, since the meaning of finite has to do with the Cauchy horizon. Since geodesics incompleteness occurs, the spacetime has definitely a Cauchy horizon, and the word finite, indicates that the union of this Cauchy horizon  of the initial hypersurface $\mathcal{S}(\tau_i)$ with this initial hypersurface $\mathcal{S}(\tau_i)$ is compact, that is, $\mathcal{H}^{+}(\mathcal{S}(\tau_i))\bigcap \mathcal{B}$ is a compact set. This means in our case that the topology change occurs ``inside'' the non-compact spacetime subset $\mathcal{B}$ bounded by the hypersurfaces $\mathcal{S}(\tau_i)$ and $\mathcal{S}'(t_f)$.

Let us discuss Tipler's conditions on a cosmological setting. The first condition, that is, the weak energy condition and the Einstein equations, for the standard Einstein Hilbert approach, is rather impossible to satisfy for ordinary matter fluids. In Einstein cosmology, ordinary matter fluids lead to different types of singularities, but for the Big Rip to occur, a phantom fluid is needed \cite{Caldwell:2003vq}. Also the weak energy condition must also be satisfied in order for the theorem to hold true. Thus, the first condition of Tipler can only be satisfied in a modified gravity setting, in which case the Einstein equations are written in a conventional form. For example in the $f(R)$ gravity case we have,
\begin{align}\label{modifiedeinsteineqns}
R_{\mu \nu}-\frac{1}{2}Rg_{\mu
\nu}=\frac{\kappa^2}{F'(R)}\Big{(}T_{\mu
\nu}+\frac{1}{\kappa}\Big{[}\frac{F(R)-RF'(R)}{2}g_{\mu
\nu}+\nabla_{\mu}\nabla_{\nu}F'(R)-g_{\mu \nu}\square
F'(R)\Big{]}\Big{)}.
\end{align}
where the effective geometrically originating energy momentum tensor $T^{eff}_{\mu
\nu}$ is equal to,
\begin{equation}\label{newenrgymom}
T^{eff}_{\mu
\nu}=\frac{1}{\kappa}\Big{[}\frac{F(R)-RF'(R)}{2}g_{\mu
\nu}+\nabla_{\mu}\nabla_{\nu}F'(R)-g_{\mu \nu}\square
F'(R)\Big{]}\, .
\end{equation}
In this case, $T^{eff}_{\mu
\nu}$ describes the contribution of geometry and do not involve any conventional matter fluid form, this is a purely geometric contribution. Thus in modified gravity, the weak energy condition can be satisfied, if the functional form of the $f(R)$ gravity is appropriately constrained, see for example \cite{Santos:2007bs}. Now the second condition in Tipler's theorem can be satisfied for any homogeneous and isotropic spacetime of any spatial section, since it describes the fact that the geodesics congruences can focus. The essential condition for the singularity occurrence is the condition 3, which simply states that the timelike geodesics threading the initial boundary partial Cauchy hypersurface $\mathcal{S}(\tau_i)$ and preceding to the intersection of the interior of the Cauchy development of $\mathcal{S}(\tau_i)$ with the non-compact subset $\mathcal{B}$, namely $[\mathrm{int}\mathcal{D}^{+}(\mathcal{S}(\tau_i))]\bigcap \mathcal{B}$, has a finite length (finite proper time integral). This means that the geodesics incompleteness occurs somewhere in $[\mathrm{int}\mathcal{D}^{+}(\mathcal{S}(\tau_i))]\bigcap \mathcal{B}$, so after the compact partial Cauchy hypersurface compact boundary of $\mathcal{B}$, namely $\mathcal{S}(\tau_i)$ and before the diffeomorphically different compact boundary $\mathcal{S}(\tau_f)=\mathcal{S}'$. This is the clue point of the theorem, which applies in the Big Rip case, the singularity actually occurs at some intermediate partial Cauchy spacelike hypersurface, $\mathcal{S}(\tau_s)$, which belongs to the Cauchy development of the initial boundary $\mathcal{S}(\tau_i)$ and the timelike geodesics terminate on this hypersurface. Then, what follows is that topology change is possible, under the above stringent constraints, and the initial topology spacelike slice of the Universe would be $\mathcal{S}(\tau_i)$ and the final would be $\mathcal{S}(\tau_f)$. The most important features of a topology change at a Big Rip singularity is that it can only occur in a modified gravity context, and the spacetime must have compact spatial sections, not necessarily positively curved. In addition, topology change occurs in a non-compact subspace $\mathcal{B}$ of the total spacetime $\mathcal{M}$, which is finite, in the meaning we discussed above ($\mathcal{H}^{+}(\mathcal{S}(\tau_i))\bigcap \mathcal{B}$ is a compact set) and also this subset $\mathcal{B}$ is bounded by the topologically distinct hypersurfaces $\mathcal{S}(\tau_i)=\mathcal{S}$ and $\mathcal{S}(\tau_f)=\mathcal{S}'$. Also it is quite important to stress that one should not think of the subset $\mathcal{B}$ to have the trivial topology $\mathcal{S}(\tau_i)\times [\tau_i,\tau_f]$, since $\mathcal{B}$ is not compact. To the non-compactness of $\mathcal{B}$ contributes the singularity hypersurface $\mathcal{S}(\tau_s)$, which cannot easily depicted schematically, and its topology is peculiar. In fact, in order to have a non-compact and finite subset $\mathcal{B}$, this can occur if the hypersurface $\mathcal{S}(\tau_s)$ is topologically distinct from $\mathcal{S}(\tau_i)$ and it actually coincides with $\mathcal{S}(\tau_f)$, so it is the boundary of $\mathcal{B}$. In addition, since $\mathcal{B}$ is non-compact, this means that if we consider a timelike curve in $\mathcal{B}$, which starts on $\mathcal{S}$, it has no future end point that can be defined formally as the limit of a sequence of points $P_i(\tau)$ parameterized by $\tau$ which starts from $\mathcal{S}$.

We need to stress that the actual mechanism of topology change in the context of classical cosmology, as in the context we described above, cannot be easily found. In addition, it is not easy to even discuss what might occur after the singular hypersurface is reached, when geodesics of our Universe fail to proceed. Our physical description ends there, and one possibility is that topology change occurs. What would the Universe look like after this change is unknown, and this is far beyond the purposes of this work. We merely examined the possibility of topology change in a classical context, and we examined what conditions should hold true in order to formally achieve this topology change. We did not address the possibility of having open boundary spacelike hypersurfaces in the subset spacetime that topology change occurs, but this is also a possibility.

\section{Discussion and Conclusions}

In this paper we critically examined the possibility that at the Big Rip the Universe may change its topology. We discussed in some detail all the necessary conditions that must hold true in order for this topology change to occur. Following Tipler's arguments, we demonstrated that this may be possible in a finite region of spacetime $\mathcal{B}$, which has not the topology $\mathcal{S}\times [0,1]$, so it has not a trivial topology, and $\mathcal{S}$ is the boundary of $\mathcal{B}$ which is a partial Cauchy hypersurface. Also, the topologically distinct boundary of the spacetime region $\mathcal{B}$ lies in the Cauchy development of $\mathcal{D}^{+}(\mathcal{S})$ and it must be the surface for which the geodesics incompleteness occurs. Also we demonstrated that this possibility of topology change can only occur for a modified gravity, since it is the only case for which the weak energy condition, vital for the topology change, can be satisfied without introducing non-standard forms of matter. One question is, does spacetime evolves normally after the topology change? This is however a formidable task to answer, by using the current theoretical descriptions. In addition, it is possible that a quantum description will alter the singularity, or even completely remove it from the theory. For example in Ref. \cite{Alonso-Serrano:2018zpi}, in a simple quantum cosmology framework, the De-Witt criterium completely removes the Big Rip from an $f(R)$ gravity. In addition, a quantum cosmological theory may cause some sort of ``blurring'' of the singularities, see for example \cite{Smith:1986tq}. Finally, in the context of modified gravity it also possible to avoid finite-time singularities \cite{Astashenok:2012tv,Capozziello:2009hc,Nojiri:2004pf}.

Coming back to the topology change issue, we are far from determining how this topology change can occur, or find in real terms what this would imply for the physical quantities on the singular and after the singular hypersurface. We only addressed the question if formally this can be possible. We did not address the topology change issue for spacetimes having non-compact spatial sections, and this issue should be formally addressed too.

\end{document}